\documentclass[10pt,conference]{IEEEtran}
\usepackage{graphicx} % Required for inserting images
\usepackage{xcolor}
\usepackage{hyperref}
\usepackage{xspace}
\usepackage[font=small,skip=0pt]{caption}
\usepackage{paralist}
\usepackage{balance}

\newcommand{\dataset}{\textsc{Myriad People}\xspace}
\newcommand{\loam}{\textsc{loam}\xspace}
\newcommand{\myriad}{\textsc{myriad}\xspace}
\newcommand{\rethread}{re\textbar thread\xspace}
\newcommand{\nbartworks}{$9$\xspace}
\title{\dataset\\ Open Source Software for New Media Arts}
\author{
\IEEEauthorblockN{Benoit Baudry} 
\IEEEauthorblockA{\textit{Universit\'e de Montr\'eal, QC, Canada}\\
benoit.baudry@umontreal.ca}
\and
\IEEEauthorblockN{Erik Natanael Gustafsson, Roni Kaufman, Maria Kling}
\IEEEauthorblockA{\textit{Independent artists, Stockholm, Sweden}\\
erik@eriknatanael.com, ronikaufman97@gmail.com, mariakling.me@gmail.com}
}
\begin{document}

\maketitle

\begin{abstract}
New media art builds on top of rich software stacks. Blending multiple media such as code, light or sound , new media artists integrate various types of software to draw, animate, control  or synchronize different parts of an artwork. Yet, the artworks rarely credit software and all the developers involved. 

In this work, we present \dataset, an original dataset of open source projects and their contributors, which span various software layers used in new media art installations. 
To collect this dataset, we released an open call for artists and eventually curated \nbartworks artworks, which use a variety of software and media. In October 2024, we organized  a collective exhibition in Stockholm, entitled \myriad, which showcased the \nbartworks artworks. The \dataset dataset includes the 124 open source projects used in one or more of the  \myriad's artworks, as well as all the contributors to these projects. In this paper, we present the dataset, as well as the possible usages of this dataset for software and art research.
\end{abstract}

\section{Introduction}

Since the early days of computer art \cite{molnar1975toward}, the importance of software has kept growing within the arts. New media art \cite{manovich2002language}, i.e. artworks produced by means of electronic media technologies, spans a variety of art practices and genres. Examples include browser art  \cite{GustafssonWB20} that use, hack, mock or subvert web browsers and the Internet culture, immersive audiovisual installations \cite{palmer2018ryoji}, which assemble large digital systems to process data,  generate abstract visuals, and synthesize soundscapes, or the niche genre of  software art \cite{cramer2001software}, which produces artworks that reveal the inherent properties of software.

Early experiments of computer art required access to rare computing and engineering expertise, which could only be provided by large institutions \cite{franco2022interview}. The Experiments in Art and Technology (E.A.T.) \cite{loewen1975experiments} is a notable example of a successful collaboration between Bell labs and conceptual artists from the 1970's. At that time software for art was very specific to each installation and hardly shared among artists. With the development of commodity hardware and software, a wider population of artists integrated software into their practices, mostly using commercial software, such as Adobe's Photoshop or Microsoft's Kinect. Eventually, open source software solutions emerged in all sectors, from graphics to signal processing and edition.  Today, some widely used open source projects are developed by artists \cite{jennings2005creativity,mclean2011artist}. Notable examples  include the Processing library \cite{reas2003processing} for visual arts, or Pure Data \cite{puckette1997pure} for sound and music.

While the adoption of, and contribution to open source software grows in new media art \cite{halonen2007open} , the collective contributions to software are often overlooked when the final artwork is exhibited. 
The notion of the singular genius artist or other inventor developed during the renaissance\cite{mcdermott2006situating} is still very influential on people's perception of creative achievements. It is harder for us to grasp the reality of large research groups, working collectives, producers and the historical and ongoing contributions of a myriad of people to any tool and amenity used.

In this work, we present 
\dataset \footnote{\dataset is available on \href{https://zenodo.org/records/14720492}{Zenodo}; reproduction scripts are on 
\href{https://github.com/rethread-studio/rethread/tree/master/code/myriad/loam_paper/dataset}{Github}}, a dataset of open source projects and their contributors, which highlights the variety of software involved in new media art, as well as the vast community of contributors who support the endless creativity of artists. 
The dataset was gathered in the context of a collective art exhibition, for which we curated \nbartworks artworks. Artworks were selected to encompass diverse technology, as well as diverse perspectives and subjective artistic quality. The \nbartworks artworks rely on a variety of media such as sound, images, video, generative algorithms, typography, as well as a variety of displays and interactions.

%The prerequisite for the open call was that the artist had used digital technology in some way in the process of creating the work. While this can be true for almost any artist today, even when the end result is purely analog, all of the 33 applicants used quite explicit digital technologies. 

% \todo{one paragraph description of the dataset} DONE

The \dataset dataset consists of 124 open source projects hosted on  GitHub, which all serve a purpose in one or more of the \nbartworks new media artworks that we curated. We manually categorize these software projects according to the role software plays in an artwork, such as graphics, embedded system or natural language processing. This software ecosystem for new media art includes brand new technology as well as projects that have been on Github since its early days. This open source software technology is  the collective achievement of thousands of contributors.

\section{Methodology}

%\todo{are there outcomes of panel 1 that we can use here?}

\subsection{Artistic research methodology}

Our objective is to collect a set of artworks that represent diverse practices and use different media. This diversity objective serves different purposes: collect diverse open source projects used for art; engage a wide audience into the conversation;  and capture the intricate nature of projects building upon projects. 
The core of our methodology relies on the organization of collective art exhibition that questions the notion of authorship in new media arts, entitled \myriad.

Through our previous experiments with open source software and art installations, as part of the \rethread collective\footnote{\url{https://rethread.art/}}, we experimented with extensive forms of crediting. Based on these past experiences we distinguish between three categories of new media art credits. First, the artist(s): who or what would conventionally be perceived as the author/creator of a given piece of art? Second, the software: which software has been used in the creation of a given piece of software art, who has contributed to creating this software and how could we trace them technically and display them legally? Third, other contributions: who or what else has contributed to the existence of a given digital art piece other than the artist or software? Examples of such contributions include inspirations or production and technical staff.

In the fall 2023, we published an open call for artists who use software in their work. To reach a wide and inclusive population of artists, the call was distributed digitally via social media and on multiple artist call websites. %\todo{would be nice to have a link to the list of channels that we used}.  
The call specified the theme of the exhibition, ``credit for contributors in art and software'',  and the expectation that selected artists would submit a comprehensive list of credits, including all involved artists, software and other contributors. With support of Kulturbryggan, part of the Swedish Arts Grants Committee, we could offer to pay the selected artists a fee of 17000 SEK (approx. 2000 euros) for exhibiting their work and pay for transport of artwork and artist. %\todo{add a footnote with link to the call}

We received 33 applications and selected \nbartworks artworks via a qualitative review process with a focus on diversity and thematic suitability.
Selected artists were interviewed, briefed on the project and then asked to provide an extensive list of collaborators, software and contributors via a prepared template. The interpretation of what "extensive" should mean in this context and who or what to include was intentionally left to each artist. This led to software lists ranging from 9 to 39 items, with 0 to 21 additional digital tools mentioned and the lists of people who contributed through other means ranging from 0 to 39 very diverse mentions.

\subsection{The curated new media artworks}

% \todo{Explain the diversity of artworks present at Myriad, which serve as a the raw material for the \dataset dataset and the \dataset installation: CNC / sand machine; interactive music generation, AI and literature, etc.} DONE

The following artworks were selected for \myriad. Following convention and practicality, this list contains the title, the artist(s), their nationality and a short description of the work. Software and other contributors are detailed in the next section where we present the  \dataset dataset. An extensive description of the artworks is available on the site of \myriad \footnote{\url{https://rethread.art/projects/myriad/}}.
    
\textit{A not so distant past} [ANSTP], by Steve Ashby (USA): a generative sound work, reinterpreting color scale data of video sources as control sources for sound creation via note selection, amplitude modulation, and other parameters.
    
\textit{Apocryph} [A], by Nicolas Boillot (FR/CH): an installation taking the form of a hybrid book, a chimera of text and images, questioning the power and ethics of generative algorithms and the authorship of AI-produced works.
    
\textit{Dear Ai} [DA], by Fred Wordie (UK): an installation showcasing a selection of algorithmically handwritten AI letters, inviting us to ask whether "I miss you" scrawled on the back of a postcard is worth more than 5000 words of AI love? 
    
\textit{Glommen} [G], by Jonas Johansson (SE): a deeply personal audiovisual work, a generative sunset reflecting on the artist's connection to his seaside hometown.
    
\textit{Infinite Sand Sorter} [ISS], by Agoston Nagy (HU): a drawing machine caught in an unending cycle of making very slow movements, echoing ancient methods of conceptualizing ideas in the sand.
    
\textit{Megatouch} [M], by Håkan Lidbo and Per-Olov Jernberg (SE): a $6\times1$ meter touch surface where visitors can discover hundreds of sounds, rhythms and animations, all perfectly synchronized.
    
\textit{Pain Creature} [PC], by nar.interactive (IT, TR): an interactive dance performance exploring the communication of bodily discomforts and the dynamics of witnessing others' pain.
    
\textit{RELAXRELAXRELAX} [R], by Yuvia Maini (SE): an interactive installation discussing the individual pursuit of "presence" in Western mindfulness culture.
    
\textit{Why Am I Seeing This?} [WAIST], by Ivana Tkalčić (HR): a video installation, exploring the influence of Information, Communication, and Technologies (ICT) on human perception and reality.

\section{The \dataset Dataset}

The artists who took part in the \myriad exhibition listed the open source projects  included in their artwork. This list of projects is the seed that we use to consolidate and preserve the \dataset dataset. 

\subsection{Collecting \dataset}

%\todo{discuss the variety of libraries, programming languages, size of projects, and the various purpose that software plays in the artworks (visuals, network, learning, DSP, etc.)} DONE

Starting from the lists of projects provided by the artists, we collected all the ones that were hosted on Github. This gave a list of 124 repositories
We did not include repositories that are not published on GitHub, since we should have treated them case by case.  Mozilla is a special case who has a "credits" page on their website listing all their contributors, "whose efforts over many years have made this software what it is"\footnote{\url{https://www.mozilla.org/credits/}}.

We fetched the contributors of each repository through the GitHub API. There are two main types of contributors: logged-in users (for whom we can get the GitHub username) and anonymous ones (who were not logged-in to GitHub when pushing the changes). Some repositories have a very high number of anonymous contributors, because they are public mirrors of repositories hosted elsewhere, for example V8. In addition, the API fails when trying to fetch repositories that are too large, namely Linux and Chromium. We were unable to obtain the data for those projects. 

% I don't know where this reflection should go
There is a stark contrast in the number of dependencies of code bases in different languages. Modern languages like Rust and JavaScript encourage the use of dependencies due to their competent package managers, whereas languages such as C and C++ tend to lead to an obfuscation of the real contribution record when code is manually copied between projects. It is safe to assume that contributors in our dataset are a subset of those connected to the exhibited works.

\subsection{The \dataset dataset}

The \dataset dataset covers 124 repositories where contributions have been made by 14797 unique logged-in contributors and 54379 anonymous ones. There open source projects have different purposes, which we have manually classified in one of the following 9 categories: 
\begin{inparaenum}[(i)]
    \item Artificial intelligence and machine learning (e.g. TensorFlow);
    \item command-line tool (e.g. NPM CLI);
    \item app/website/program with a GUI (e.g. GIMP);
    \item related to electronics (e.g. Arduino IDE);
    \item engine (e.g. V8);
    \item programming language/library/plugin specifically to create/process something visual (e.g. openFrameworks);
    \item programming language/library/plugin specifically to create/process sound or music (e.g. Pure Data);
    \item text and speech (e.g. Festvox);
    \item other (e.g. Python);
\end{inparaenum}
When a project  fits  more than one category, we choose the most specific one.

\begin{figure}[th]
    \includegraphics[width=0.9\columnwidth]{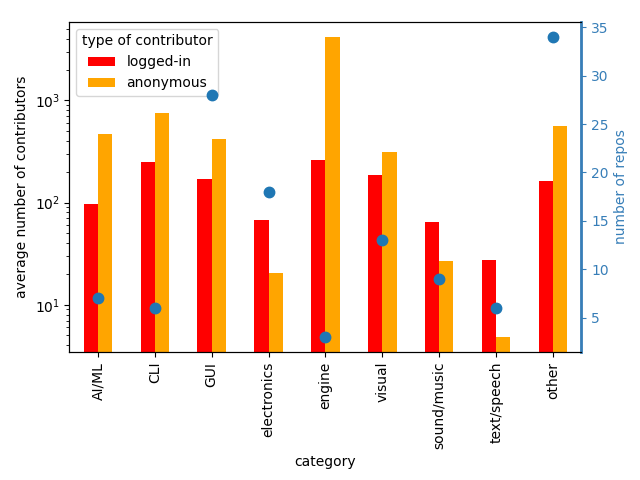}   
    \caption{Average number of contributors and number of repositories, per category}
    \centering
    \label{fig:contributors}
\end{figure}

The repository with the largest number of contributions is Gecko, from Mozilla, with 937482 contributions. According to the statistics on the GitHub website, Chromium and Linux, which we were not able to fetch through the API, have a history of more than, respectively, 15 and 13 millions commits, at the time of writing. The one with the least amount of contributions is ElizaBot, a JavaScript bot based on Joseph Weizenbaum's ELIZA \cite{Weizenbaum1966ELIZA}, with only 4 contributions, all done by the same user and on the same date, in 2012. Creative coding libraries, such as p5.js, created by an artist (Lauren Lee McCarthy) and maintained by other artists, typically have a tens of thousands of contributions.

\begin{figure}[th]
    \includegraphics[width=0.9\columnwidth]{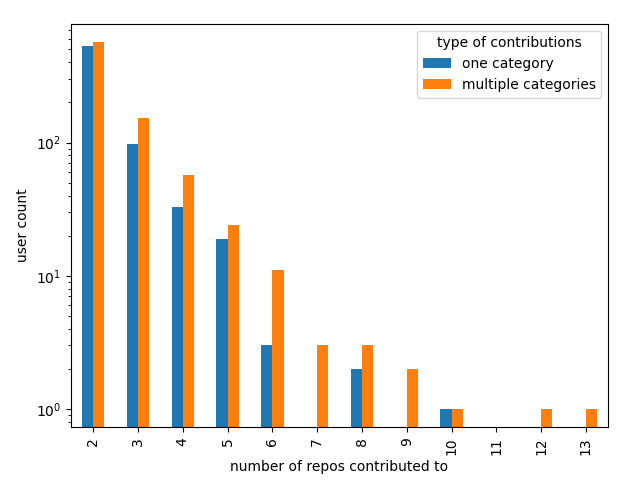}   
    \caption{Users contributing to multiple projects (no bots)}
    \centering
    \label{fig:multi-contributors}
\end{figure}

\autoref{fig:multi-contributors} focuses on the human developers who contribute to more than one project in \dataset. We distinguish between developers who contribute to multiple projects in the same category, e.g. different sound generation libraries, and developers who contribute to different categories. 

The most prolific (non-bot) contributor in our dataset, is eltociear, who contributed to 13 different projects, across 7 different categories, from "command-line tool" to  "programming language/library/plugin specifically to create/process something visual". Then, dtolnay contributed to 12 different repositories, including 5 projects as an owner. Indeed, dtolnay is involved with the official Rust standard library team, and personally writes and maintains several key libraries in the Rust ecosystem. The third- to sixth-biggest contributors were also mostly involved in those Rust repositories.

\begin{figure}[th]
    \includegraphics[width=0.9\columnwidth]{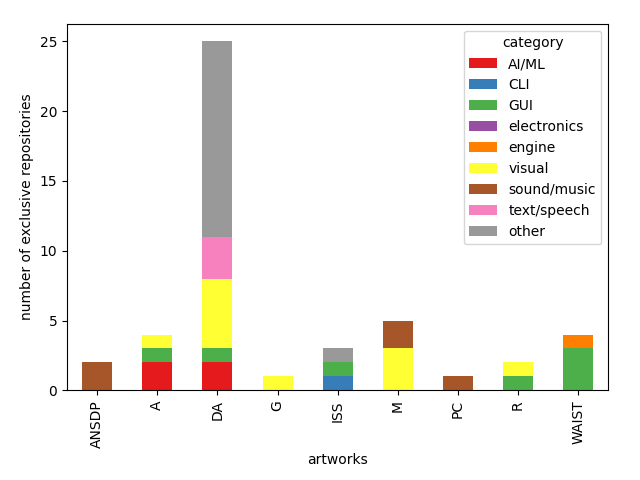}   
    \caption{Number of repositories exclusively used in each artwork}
    \centering
    \label{fig:unique-projects}
\end{figure}

An important goal of our curation process was to gather artworks that represent diverse practices, to capture a comprehensive set of open source projects used for the arts. \autoref{fig:unique-projects} shows the number of projects that were used by one single artwork. We see that all artworks used at least one project that no other work used. For example, \emph{Glommen} (G) was the only work using Three.js, or \emph{A Not So Distant Past} (ANSDP) was the only work using Envelop For Live and the IEM Plug-in Suite. It is interesting to note that \emph{Dear Ai} (DA) used a very specific software stack, which included 25 projects used exclusively for this work. Indeed, the project consists of multiple different components, from a website to the interactive installation. In particular, it used scikit-learn, Sass and Plottie, a plotting/cutting program for Silhouette Plotters.

%   Regarding future work with this data analysis and visualization, one could explore clustering contributors and seeing if the formed communities match with the categories we defined here.

\section{Possible usages of \dataset}

In this section we first explain how we closed the loop of the \myriad exhibition, building a specific art installation that is fed by the  \dataset dataset. Then, we discuss how the dataset can be used for future research in software engineering. 

\subsection{The \loam art installation}

% Programming languages used directly: SuperCollider, JS, Rust, C++, Python, (Excel style forumlas)
In 2024, we designed and built the \loam audiovisual art installation, which is fed by the \dataset dataset. \loam consists of three parts letting the audience make sense of the vast communities of developers and software involved in the operation of the artworks presented in the \myriad exhibition. The exhibition took place in October 2024 in Reaktorhallen (R1) in Stockholm. The location is a former nuclear research reactor and the central sculpture was mounted in the same space where the reaction used to take place.

\begin{figure}
    \centering
    \includegraphics[width=1\linewidth]{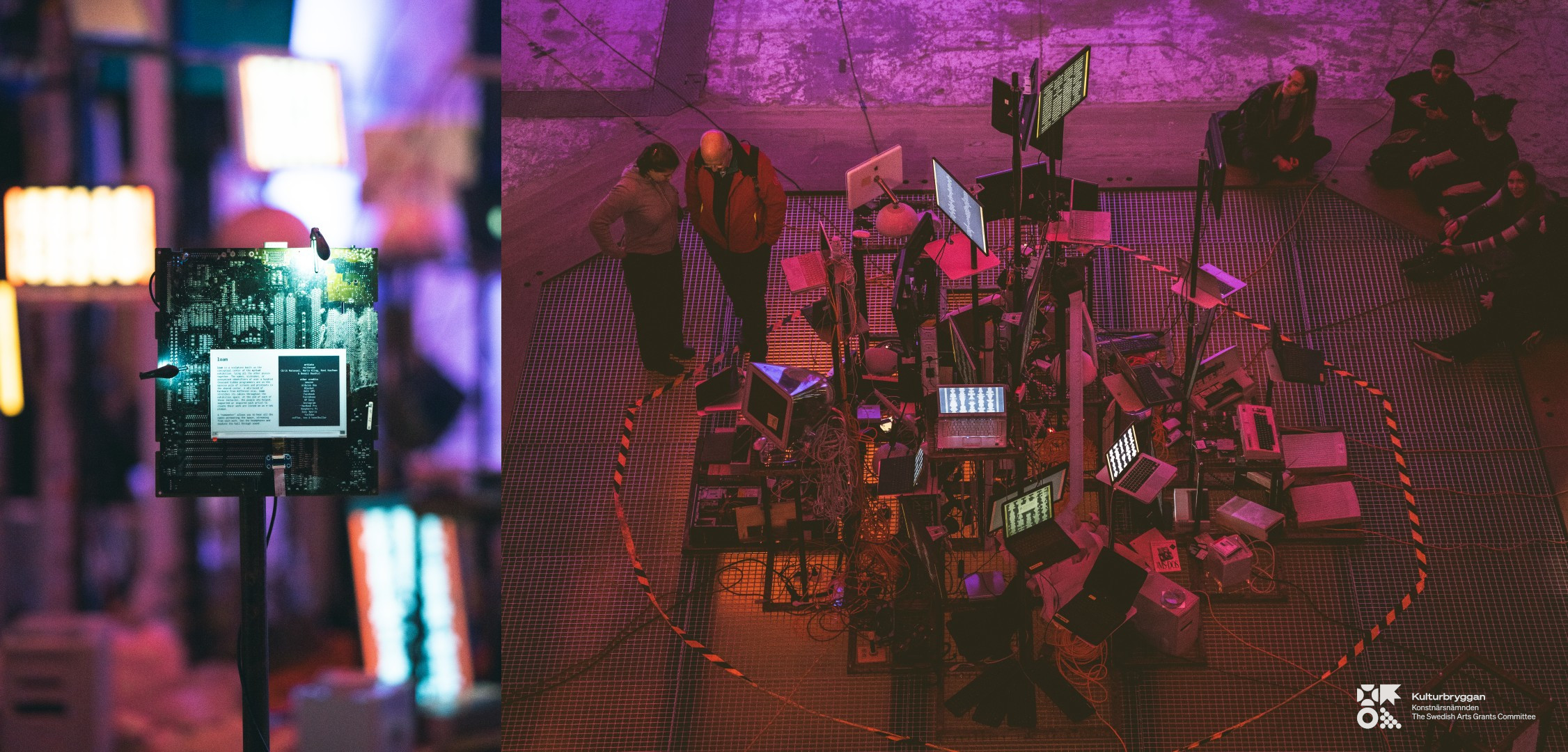}
    \caption{The \loam sculpture is an art installation that is fueled by the \dataset dataset. The photo on the right shows the central \loam sculpture from above, displaying thousands of contributor names on historical computers; the photo on the left is an e-ink display showing non-developer contributors for an artwork.}
    \label{fig:loam}
\end{figure}

% Erik: I think mentioning programming languages is relevant to the dependencies?
The first part of \loam, is a sculpture that is in the center of the \myriad exhibition. It is shown on the right side of \autoref{fig:loam}. It displays the names of all software contributors in \dataset.  It is made with  recycled and historical computers to represent the parallel development of hardware and software through time, running up to 20 years old versions of macOS and Windows. The sculpture was mounted into a  $3m \times 4m \times 4m$ structure, with cables, and electronics creating an intentionally messy network thicket.  
The screen based visualizations show scrolling lists of tens of thousands of open source contributor usernames.% or hashed email addresses for anonymous contributors.%, organized by repository.  Closed software from the lists, such as Adobe or Windows, were represented via still images with the software name and representations of redacted names.
%Some large and fundamental repositories were printed out using a thermal printer during the exhibition with the names gathering in an increasing heap within the sculpture, serving to emphasize the mass of contributors as well as as a reference to Margaret Hamilton's print out of the Apollo 11 code. 

The second part consists of e-ink displays by the side of each artwork. They show names of non-developers contributors. The aesthetics of these modules was a play on the classical convention of plaques in museums. The cables from the central sculpture to each module tied the whole \loam installation together, emphasizing how every work is always connected to a wider cultural and technological context. %, help the audience to orient themselves in the exhibition and provide the electricity for the e-ink displays and the Daisy Seed for the specialized sound work.
The third part of \loam is a sonic device, called a \textsc{nomometer}. It was developed for \dataset. It is a spatially sensitive sonic device which, as you get close to each work in the exhibition, plays back the names of any known contributor to any layer of its digital creation. Unavailable names are replaced by noise.

%The software used in and for \loam is not included in \dataset, but this is a short overview: (A) uses p5.js running in Chromium on Raspberry Pi OS on Raspberry Pi 5 hardware. The "modules" (B) are using micro-controllers programmed in Arduino C++ based on the manufacturer's example code, and a small utility program in Rust used to convert image frames into bit encoded byte arrays. (C) uses software for sound of two kinds: music composition/DSP, and playback infrastructure. Voice material for the piece was first generated using Coqui TTS and mimic3. The work was composed in SuperCollider, a programming language and environment for sound and music. To play back the piece, a custom hardware device was constructed consisting of a Daisy Seed and a DW3000 UWB module connected to an ESP32, communicating via SPI. The ESP32 was programmed in Arduino C++ with few additional dependencies. The Daisy Seed was programmed in Rust, largely using the HAL\footnote{Hardware Abstraction Layer} crates, which makes for a large dependency tree, but with many authors shared among dependencies.

\subsection{Further Usages in Software Research}

%open source software and art
Art is overlooked as an application domain for software research. The \dataset dataset serves as a starting point to further explore the different intersections that exist between new media art and software engineering. 

%the role of contributors in open source for art
\emph{Open source contributors.} The software engineering community has devoted significant effort in understanding the motivations and socio-technical dynamics among open source contributors. For example, previous work has analyzed the role of these contributors in the areas of scientific software  \cite{Milewicz19} or postgresql \cite{German06}, and our dataset can consolidate this body of knowledge with open source contributors for new media arts. The p5.js community of systematically acts to maintain a diverse and inclusive population of developers and contributors. \dataset can serve to further research onboarding practices  \cite{foundjem2021open}, turnover \cite{jamieson2024predicting} or gender representations \cite{qiu2023gender} in artistic communities in comparison to other domains.

\emph{Computer science education.} Creative coding and digital arts are attractive approaches to introduce a diverse audience to programming and computer science \cite{xu2018updating}. This approach makes programming enjoyable to beginners \cite{balter2010enjoying} or supports the introduction of programming concepts to non-programmers \cite{terroso2022programming}. Approaching computer science through the lens of the arts is also a way to let computer scientists embrace a broader perspective on the aesthetics of programming,  moving away from the utilitarian aspect of software \cite{bond2005software}. \dataset provides a rich source of inspiration for future educators who wish to include creative coding in courses that rely on different programming languages (from javascript to rust).

\emph{Programming abstractions for new media arts.} The end-to-end design and deployment of new media art installations relies on multiple levels of programming abstractions \cite{roberts2014gibber}. For example, programming APIs for live coding music \cite{mclean2014making} are different from the features needed for controlling a drawing \cite{baudry2024programming} or knitting \cite{subbaraman2022p5} machine. Programming models can also be tailored to art installations, for example to optimize energy to color display \cite{stanley2016crayon} or to automatically analyze music code \cite{gold2011}. 
With its wide diversity of programming languages, the \dataset dataset is an opportunity to build an inclusive body of knowledge about  APIs or performance considerations when building software for the arts.

%Use this dataset for preservation; Software-based art \cite{falcao2019preservation}; 
%\href{https://sourcegraph.com/blog/strange-loop/strange-loop-2019-recreating-forgotten-programming-languages-for-art}{Strange Loop 2019 - Recreating forgotten programming languages, for art!}; 
%\url{https://github.com/thi-ng}

\section{Conclusion}

We have collected an original dataset of 124 open source repositories and their contributors, which form the foundations of a diverse set of \nbartworks new media art installations. The dataset is called \dataset.
With this work we emphasize the role of software and  developers within the arts world. Future works include studying how open source software contributes to the success of new media artists \cite{fraiberger2018quantifying} and how esolangs enchant the arts \cite{temkin2017language}.

\balance
\bibliographystyle{IEEEtran}
\bibliography{artbiblio.bib}

\end{document}